# Gouy-phase-mediated propagation variations and revivals of transverse structure in vectorially structured light

Ru-Yue Zhong,[1] Zhi-Han Zhu,[1,*] Hai-Jun Wu,[1] Carmelo Rosales-Guzmán,[1,2] Shu-Wei Song,[1,†] and Bao-Sen Shi[1,3]

[1] *Wang Da-Heng Center, Heilongjiang Provincial Key Laboratory of Quantum Control,*
*Harbin University of Science and Technology, Harbin 150080, China*
[2] *Centro de Investigaciones en Óptica, A.C., Loma del Bosque 115, Colonia Lomas del campestre, 37150 León, Gto., Mexico*
[3] *CAS Key Laboratory of Quantum Information, University of Science and Technology of China, Hefei, 230026, China*

(Compiled 30 April 2020)

Exploring the physics and potential applications of vectorially structured light with propagation-invariant transverse structures has benefited many areas of modern optics and photonics. In this paper, we investigate the non-eigen vector modes of paraxial light fields, focusing on the propagation variations and revivals of their transverse structures, including both spatial and polarization structures. We show that the physical mechanism behind the variations and revivals of their transverse structure is linked to the evolution of the intramodal phases between the constituting spatial modes. Such evolution originates from fractional Gouy phases, or rather, Geometric-phase difference between spatial modes with different orders under a same unitary transformation. This underlying principle, provides a general guideline for shaping vectorially structured light with custom propagation-evolution properties, and may also inspire a wide variety of new applications based on structured light.

## I. Introduction

Paraxial light fields with specific transverse structures, including both their complex amplitude and polarization, are generally known as structured Gaussian modes or structured light for short, which have become a very active topic in modern optics. [1-3] The transverse structure of a paraxial light field and its corresponding evolution upon propagation can fully be described by a solution of the paraxial wave equation, which can be regarded as an analogue of the Schrodinger equation for free particles [4]. Except for an ideal plane wave, any eigen solution of the wave equation corresponds to a physically realizable 'eigen' spatial mode, whose transverse structures are propagation-invariant apart from an overall change of size. For instance, the Laguerre–Gauss (LG) and the Hermite–Gauss (HG) modes, well-known in laser physics [5], are eigen solutions of the wave equation in cylindrical and Cartesian coordinates, respectively; similarly, the Ince–Gauss (IG) modes are analogs in elliptical coordinates [6]. These Gaussian-mode families have their own special features in transverse structures (but have the same fundamental $TEM_{00}$ mode), and each can independently constitute a unique and unbounded 2-D Hilbert space. Thus, in principle, one can define other new Gaussian-mode families on-demand by designing a series of coherent superpositions (i.e., complex series) of LG, HG or other accessible eigen modes with identical modal orders [7], such as the intermediate Hermite–Laguerre–Gauss modes [8]. Moreover, due to the vector nature of light, the superposition can also be realized in a vector form and to generate non-separable superpositions of orthogonal polarizations and spatial modes, which are known as optical spin-orbit coupled (SOC) modes [9]. These special vector superpositions with well-defined modal orders can, in fact, be regarded as eigen solutions of vector wave equation having features with spatially variant polarization, such as cylindrical vector modes [10].

Beyond the (scalar or vector) eigen modes, more general cases are non-eigen modes with non-identical modal orders, whose transverse structures would no longer remain constant upon free-space (or focusing) propagation. Importantly, the propagation variations of the transverse structures in these cases provides the physical basis required to build 3-D structured light [1, 11]. The 3D-tailored paraxial light in amplitude and polarization will provide the means to shape and control (both linear and nonlinear) light-matter interactions [12-15]. Noteworthy, the propagation evolution of non-eigen structured light is usually regular and even sometimes displays periodicity. In the scalar regime, for instance, Courtial *et al.* reported that the patterns of second-harmonic waves of some high-order LG modes varied upon propagation but can finally reproduce their original structures in the far field [16]; and this self-imaging in the far field was attributable to a varied intramodal phase induced by nonsynchronous Gouy-phase accumulation [17]. Recently, Khoury's group discussed, in a more general way, the pattern (intensity profile) variations and revivals of scalar structured Gaussian modes, which were enabled by the fractional Gouy phase [18,19]. Importantly, they unified the physics behind the pattern revivals and the well-known Talbot effect [20]. In the vector regime, the generation and application of vector structured light with custom propagation-variant transverse

* zhuzhihan@hrbust.edu.cn
† ssw@hrbust.edu.cn

structures have also attracted increasing interest in recent years [21-25].

In this paper, a systematic study on the evolution of the transverse structure of vector structured light upon free-space (or focusing) propagation is presented. Particularly, the intramodal-phase variations and revivals in SOC space and spatial-mode subspace mediated by the fractional Gouy phase are revealed. In addition, a detailed analysis of the associated influence on the transverse structure of vector structured light is given. The structure of the paper is as follows: a theoretical background and experimental methods are presented in Section II and following this we proceed to analyzed the intramodal-phase variation (IPV) and the intramodal-phase revival (IPR) induced propagation variations and revivals of vector profiles in various vector structed light in Section III. Finally, in Section IV we provide a conclusion and final remarks.

## II. Theoretical and Experimental Methods

*Theory*. — Through this manuscript, we will use the term vector structured light to refer to a non-separable superposition state with respect to a pair of orthogonal polarizations $\hat{e}_{\pm}$ and associated polarization-dependent (scalar) spatial modes $\psi_{\pm}(\mathbf{r},z)$, which at the beam waist ($z=0$) can be mathematically expressed as

$$\mathbf{E}(\mathbf{r},z_0) = \sqrt{\alpha}\psi_{+}(\mathbf{r},z_0)\hat{e}_{+} + e^{i\theta}\sqrt{1-\alpha}\psi_{-}(\mathbf{r},z_0)\hat{e}_{-},$$
$$\psi_{\pm}(\mathbf{r},z_0) = u_{\pm}(\mathbf{r},z_0)e^{iv_{\pm}(\mathbf{r},z_0)} \quad (1)$$

where $\alpha \in [0,1]$ and $\theta$ are the weight coefficient and initial intramodal phase, respectively, and $\mathbf{r}$ denotes the transverse coordinates. Given that $\psi_{\pm}(\mathbf{r},z)$ are orthogonal to each other, if $\alpha \neq 0$ or $1$, $\mathbf{E}(\mathbf{r},z_0)$ describes a spatially variant polarization state with an intensity profile of $\alpha u_{+}^2(\mathbf{r},z_0) + (1-\alpha)u_{-}^2(\mathbf{r},z_0)$. Except for some special cases, such as cylindrical vector modes, the transverse structure (or vector profile) of $\mathbf{E}(\mathbf{r},z_0)$ usually cannot remain constant upon propagation ($z > 0$). The focus of this study is to reveal underlying principle of the transvers-structure evolution of $\mathbf{E}(\mathbf{r},z_0)$ from $z_0$ to $z_\infty$.

Unlike an infinite plane wave $e^{ikz}$ having a pure on-axis wave vector $k_z = k = \omega/c$, according to the uncertainty principle concerning the transverse position and momentum, any paraxial light field with a finite aperture inevitably has a transverse-momentum component given by $k_\mathbf{r}^2 + k_z^2 = k^2$. As a result, since $c^2 = v_g v_{ph}$, the group velocity of paraxial light along the $z$ direction should be slower than $c$, while the propagation of the wavefront should be ahead of an ideal plane wave, leading to a pair of complementary expressions [26-29]

$$\tau(z) = -\int_0^z \left(\langle k_\mathbf{r}^2 \rangle / 2k^2\right) dz,$$
$$\phi(z) = \int_0^z \left(\langle k_\mathbf{r}^2 \rangle / k\right) dz \quad (2)$$

where $\tau(z)$ and $\phi(z)$ represent the group-velocity delay and the wavefront acceleration of a paraxial wave compared with an ideal plane wave, respectively, and the expected value can be calculated by using the transverse Laplacian operator, i.e., $\langle k_\mathbf{r}^2 \rangle = -\langle \psi | \nabla_\mathbf{r}^2 | \psi \rangle$. The two complementary terms co-determine the group and phase velocities of the paraxial Gaussian modes. Specifically, the first term governs the subluminal-propagation features of paraxial light and, recently, the arrival delays of various structured Gaussian pulses and photons have been successfully observed [26-28]. The second expression gives rise to the well-known Gouy-shift effect, which was first observed by Gouy and has been overserved in various wave systems [30-37]. Noteworthy, both $\tau(z)$ and $\phi(z)$ may make a profound impact on the transverse structure of the paraxial light fields via mediating temporal-spatial-overlap and intramodal-phase structures, respectively. However, in this study we only consider monochromatic waves and thus focusing on the IPV and IPR mediated by Gouy phase.

For a general propagating wave, the accumulated Gouy phase at a given propagation plane can be obtained via Kirchhoff diffraction integral (or the Collins version for a lens system), contributed by the phase factor in front of the integral [29]; while for a propagation-invariant structured Gaussian mode with a well-defined modal order $N$, there is a concise and elegant expression developed by [5]

$$\phi_N(z) = (N+1)\arctan(z/z_R), \quad (3)$$

where $z_R = kw_0^2/2$ denotes the Rayleigh distance. Based on this, we analyze the propagation evolution of $\mathbf{E}(\mathbf{r},z)$. The modal order $N$ is defined as $2p + |\ell|$ and $m+n$ for LG and HG modes, respectively. Note that the profiles of $u_{\pm}(\mathbf{r},z)$ may be propagation variant, that is, they may be superpositions of eigen modes with different modal orders. For simplicity, we first assume $\psi_{\pm}(\mathbf{r},z)$ represents a pair of propagation-invariant eigen modes with identical modal orders of $N_{\pm}$, respectively. Even under this simplified assumption, however, the initial transverse structure of $\mathbf{E}(\mathbf{r},z_0)$ usually still cannot remain constant after leaving the beam waist plane. More specifically, according to Eqs. (1) and (3), we have

$$\mathbf{E}(\mathbf{r},z) = \sqrt{\alpha}\psi_{+}(\mathbf{r},z)\hat{e}_{+} + e^{i[\theta+\delta(z)]}\sqrt{1-\alpha}\psi_{-}(\mathbf{r},z)\hat{e}_{-},$$
$$\delta(z) = \Delta N \arctan(z/z_R) \quad (4)$$

where $\Delta N = N_{+} - N_{-}$ can be both positive or negative integers. Note that the newly appeared $z$-dependent exponential term, that is, $\delta(z)$, gives rise to a propagation-varied effective intramodal phase $[\theta + \delta(z)]$. Thus, for $\Delta N \neq 0$, the effective intramodal phase will no longer remain constant since $\psi_{\pm}(\mathbf{r},z)$ carry different modal orders. This IPV phenomenon occurring in SOC space leads to a variation in the transverse polarization profiles as a function of $z$. However, it is worth noting that, first, the intensity profile given by $\alpha u_{+}^2(\mathbf{r},z) + (1-\alpha)u_{-}^2(\mathbf{r},z)$ remains constant apart from a $\sqrt{2}$ overall enlarging per $z_R$, and second, the spatial 'concurrence', given by $C = \sqrt{1-\overline{s}_3(\mathbf{r})}$, is still unchanged. Here $\overline{s}_3(\mathbf{r})$ denotes the average of the third Stokes parameter, defined by $\hat{e}_{\pm}$, over the transverse plane.

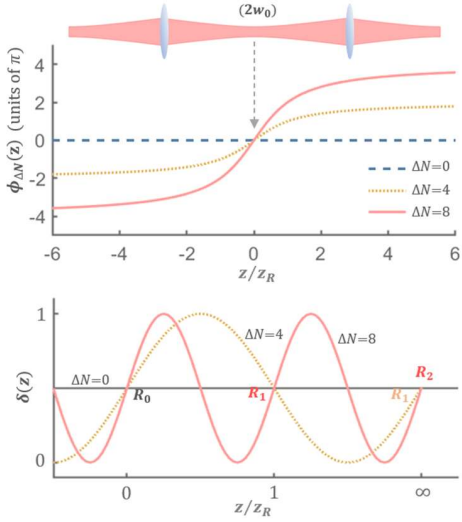

FIG. 1. Normalized Gouy-phase difference $\phi_{\Delta N}(z)$ and $\delta(z)$ as functions of propagation distance, for cases $\Delta N = 0$, 4, and 8, respectively, where $R_n$ denotes the nth IPR point.

Figure 1 shows $\delta(z)$ as a function of the propagation distance, for the cases $\Delta N = 0$, 4, and 8, respectively. It is important to note that the IPV is a periodic function, or rather, $\theta = [\theta + 2\pi n]$, $n \in Z$, thus the initial intramodal phase in SOC space is reproduced as $\delta(z) = 2\pi n$. Due to the fact that $\arctan(z/z_R) \in [0, \pi/2]$, the precondition for IPR is $\Delta N \geq 4$. This IPR phenomenon leads to revivals of the polarization profile; in other words, it provides a mechanism to realize a periodic 3-D paraxial vector structure.

In addition, usually $\psi_\pm(\mathbf{r}, z)$ may be non-eigen modes composed by $n$ eigen modes with different orders $N_1$, $N_2$, …, $N_n$, leading to their beam profiles, i.e., $u_\pm^2(\mathbf{r}, z)$, being not propagation-invariant and intensity-profile revivals may occur [18]. In these complicated cases, where the IPVs and the IPRs occur in both SOC space and spatial-mode subspace, the vector profile of $\mathbf{E}(\mathbf{r}, z)$ would experience a more intense evolution with respect to both polarization and intensity profiles, as well as the spatial 'concurrence' or $\bar{s}_3(\mathbf{r})$ defined by $\hat{e}_\pm$, being change upon propagation. These complicated cases are discussed later through specific examples.

*Experimental Setup.* — Before considering the detailed analysis based on specific vector modes, the experimental setup used is discussed briefly to verify the theoretical analysis. In the experiments, a compact and robust device was used for the generation and manipulation of vectorially structured light [38]. Figure 2 shows a schematic representation of the device, where the core component is the self-stable polarization Mach-Zehnder interferometer. To generate a desired vector mode at the beam waist plane, that is, $\mathbf{E}(\mathbf{r}, z_0)$, a laser beam (780 nm) with the TEM$_{00}$ mode enters the Mach-Zehnder interferometer from the left polarizing beam displacer (BD-1). After that, the two parallel beams were directed simultaneously to two different sections of a spatial light modulator (SLM) and were independently controlled. Two complex-amplitude-modulation holograms designed for producing $u_\pm(\mathbf{r}, z_0)$ exactly were loaded on the two sections [39]. Then, another polarizing beam displacer (BD-2) combined $u_\pm(\mathbf{r}, z_0)$, and outputted the desired vector mode $\mathbf{E}(\mathbf{r}, z_0)$. To characterize the transverse structure of the vector structured light, a camera (CCD) combined with polarizers was used to perform spatial Stokes tomography on the generated modes [40]. Besides, to realize fine control of the propagation distance, that is, $z/z_R$, the digital propagation phases were superposed on the holograms [39]. In this way the camera fixed at the back-focal plane of the Fourier lens (FL) can record exactly the 3-D vector structures of incident light without any mechanical movement [41].

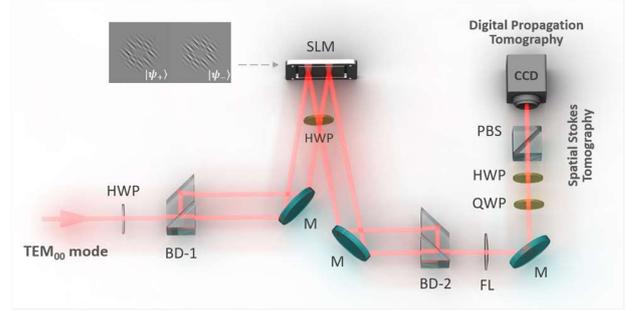

FIG. 2. Schematic representation of the experimental setup and where the key components include the polarizing beam displacer prisms (BD), a spatial light modulator (SLM), a camera (CCD), a quarter-wave plate (QWP), half-wave plates (HWP), and mirrors (M).

### III. Results

In this section, various types of variations and revivals for the specific vector structured light are further examined, which are categorized by evolution types, i.e., full propagation invariant, only polarization profile variant, and full vector profile variant, respectively. Without loss of generality, we can represent and analyze structured Gaussian modes in cylindrical coordinates, i.e., $\mathbf{r} \to \{r, \varphi\}$, and the associated eigen modes carrying orbital angular momentum (OAM) are denoted as $LG_p^{\pm \ell}$, where $\ell$ and $p$ are the azimuthal and radial indices, respectively, and the mode order is defined as $N = 2p + |\ell|$.

*Propagation Invariant.* — Before studying the various transverse evolutions upon propagation, we first examine the principle for design and generation of propagation-invariant vector modes, which, in theory, would correspond to certain eigen solutions of vector wave equation in specific coordinates. By this principle we consider a general 4-order vector mode in elliptic coordinates, and given by [42]

$$[IG_{44}^e(\mathbf{r}, z; \varepsilon) \hat{e}_H + IG_{44}^o(\mathbf{r}, z; \varepsilon) \hat{e}_V]/\sqrt{2}, \qquad (5)$$

where $e(o)$ denotes even (odd) parity, and $\varepsilon \in [0, \infty)$ defines the ellipticity of the coordinates. In particular, IG modes become the LG and HG modes with same the orders as $\varepsilon = 0$ and $\varepsilon \to \infty$ [6], respectively, in which the vector mode shown in Eq. (5) is transformed into

$$[LG_0^{-4}(\mathbf{r},z)\hat{e}_H + iLG_0^{+4}(\mathbf{r},z)\hat{e}_V]/\sqrt{2}$$
$$[HG_{40}(\mathbf{r},z)\hat{e}_L + HG_{31}(\mathbf{r},z)\hat{e}_R]/\sqrt{2} \quad . \quad (6)$$

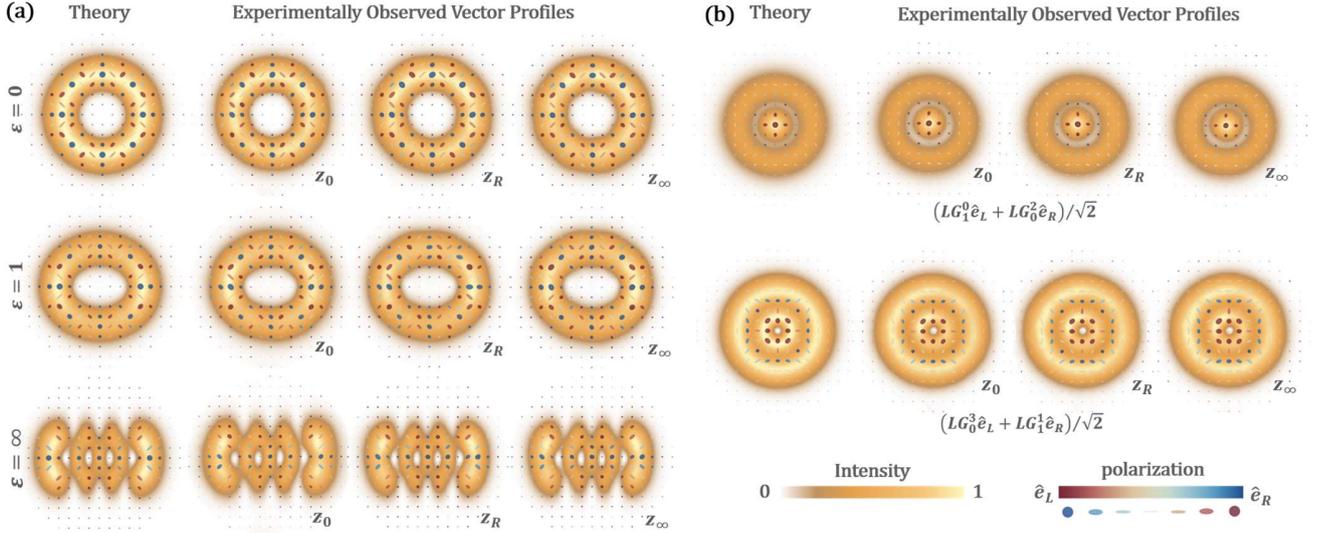

FIG. 3. General propagation-invariant vector structured Gaussian modes, where (a) shows the theoretical and observed vector profiles of 4-order general vector mode in elliptic coordinates with different ellipticity, and (b) shows two examples of propagation-invariant FP modes, the transverse unit is given in $r/w_z$.

It can be seen that all vector modes shown in Eqs. (5) and (6) have a same modal order of $N = 4$, and their theoretical vector profiles and the corresponding observations from the beam waist plane to the far field are shown in Fig. 3(a), which confirmed that the transverse structures are propagation invariant. Notes that both the HG and IG modes can be represented as a coherent superposition of a series of LG modes with the same order; for instance, the IG modes in Eq. (5), as $\varepsilon = 1$, can be represented as [6, 43]

$$IG_{44}^e = 0.036 LG_2^0 - 0.17 LG_1^{\pm 2} + 0.985 LG_0^{\pm 4}$$
$$IG_{44}^o = -0.16 LG_1^{\pm 2} + 0.987 LG_0^{\pm 4} \quad , \quad (7)$$

where one can note that all the LG components have an identical modal order, given by $N = 2p + |\ell| = 4$. Thus, besides exploring eigen solutions of vector wave equation, a more straightforward way to design a propagation-invariant vector mode is: define on-demand a pair of orthogonal spatial modes that are composed of LG (or other accessible eigen) modes with the same modal order, that is, making $\Delta N = 0$ in both SOC space and spatial-mode subspace. For instance, full Poincaré (FP) modes are well-known for their unique and propagation-rotational polarization profiles [44], and according to above principle it is possible to realize a propagation-invariant FP mode. Figure 3(b) shows two typical examples $(LG_1^0 \hat{e}_L + LG_0^2 \hat{e}_R)/\sqrt{2}$ and $(LG_0^3 \hat{e}_L + LG_1^1 \hat{e}_R)/\sqrt{2}$.

*Variation in Polarization-Profile.* — We now demonstrate the simplified case described by Eq. (4), that is, assuming $\psi_\pm(\mathbf{r},z)$ are a pair of propagation-invariant eigen modes with different orders of $N_\pm$ (i.e., $\Delta N = N_+ - N_- \neq 0$); and, moreover, for simplicity, all the initial intramodal phases were set as $\theta = 0$. For the two modal indices of the LG modes, it was assumed that $\psi_\pm(\mathbf{r},z)$ carry different azimuthal indices and modal orders, i.e., $\ell_1 \neq \ell_2$. Specifically, the FP mode, $(LG_1^0 \hat{e}_L + LG_0^2 \hat{e}_R)/\sqrt{2}$, whose polarization profile would rotate $180°$ from $z_0$ to $z_\infty$ was first considered. In Fig. 2(b) it is shown that, by adding radial mode into the lower-order mode ($LG_0^0$), its polarization profile can become propagation invariant as making $\Delta N = 0$. Here we focus on the IPV and IPR phenomena in the cases where $\Delta N \geq 4$, and we consider an FP mode $(LG_0^0 \hat{e}_L + LG_2^0 \hat{e}_R)/\sqrt{2}$, whose $\delta(z)$ and the corresponding vector profiles upon propagation are shown in Fig. 4(a). According to the simulation, the IPR phenomenon and the associated polarization-profile revival should occur at $z = 1.732 z_R$, which were confirmed by experimental observations shown in the second line. Then, the cylindrical vector mode, given by $(LG_2^1 \hat{e}_L + LG_0^{-1} \hat{e}_R)/\sqrt{2}$, was considered, as shown in Fig. 4(b). Compared with a standard radial-polarized mode, its polarization profile (inner ring) shows an oscillation from the radial (at $z_0$) to the azimuthal (at $z_R$) and finally revive the radial polarizations (at $z_\infty$). These propagation-rotation cylindrical vector modes have also been observed in the vector Bessel-Gauss beams [21], and the principle is the same: that is, making $\Delta N \geq 4$ between the two OAM conjugated modes. Now it is assumed that $\psi_\pm(\mathbf{r},z)$ carries the same azimuthal index (or OAM per photon) but a different radial index, i.e., $p_1 \neq p_2$, and we consider a such type mode with $\Delta N = 8$, given by $(LG_0^0 \hat{e}_L + LG_4^0 \hat{e}_R)/\sqrt{2}$. Figure 4(c) shows the corresponding theoretical and experimental results, as expected, two IPRs and associated polarization-profile revivals occur at $z_R$ and $z_\infty$, respectively. Unlike the two previous examples shown in Figs. 4(a) and (b), whose polarization profiles varied along azimuthal direction, here the variation occurs in a radial direction. Additionally, the intensity profiles in all the above examples have not changed, because the IPVs and the IPRs only occurred in SOC space. That is, all variations in the vector-profile originate from the rotation of the local polarization orientation, while the $C$ or $\bar{s}_3(\mathbf{r})$ of each location (or pixel on CCD) has not changed.

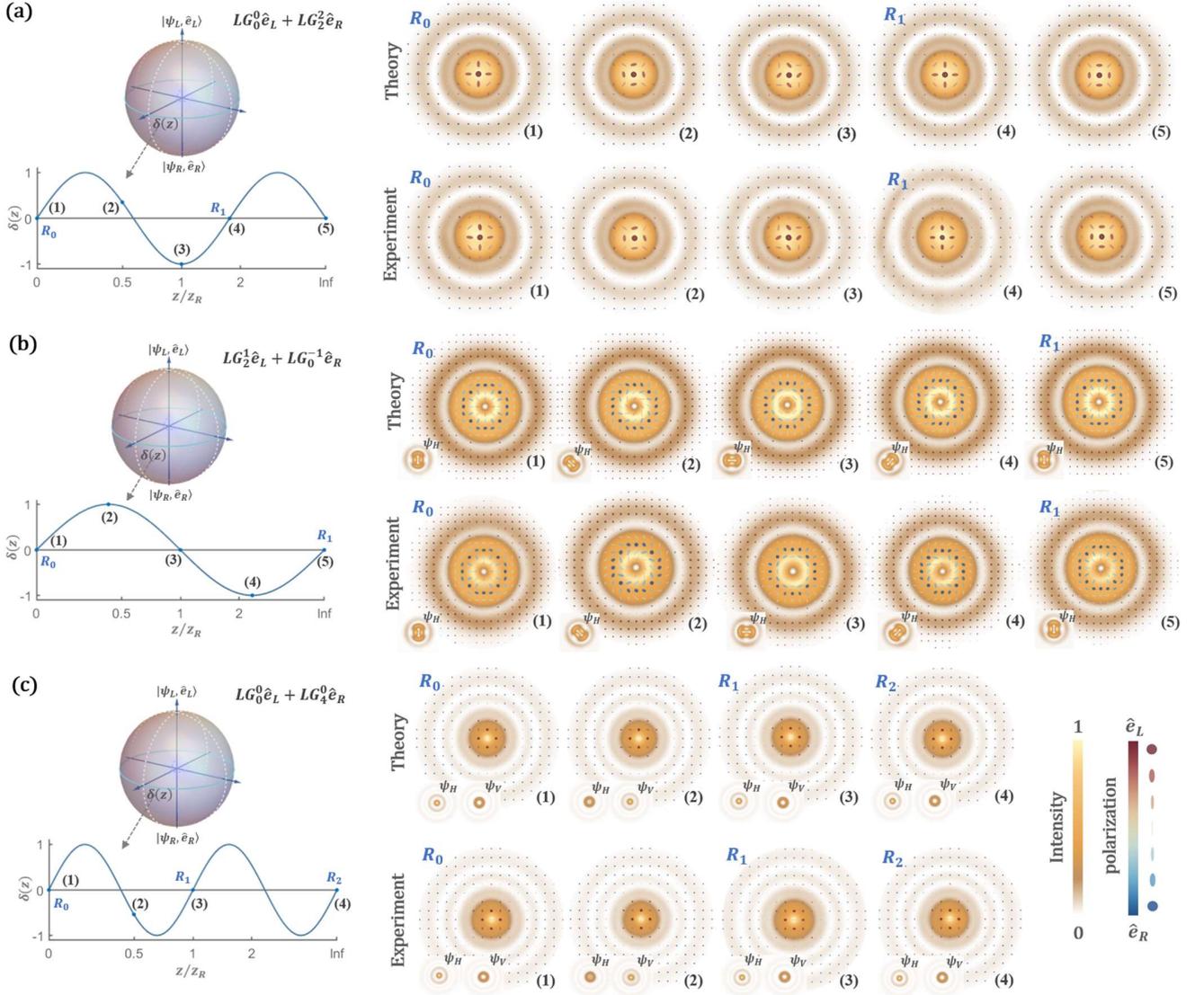

FIG. 4. Vector structured modes with propagation-variant polarization profiles, where $R_n$ denotes nth. IPR and polarization-profile revival. The transverse unit is given in $r/w_z$.

*Variation of Vector-Profile.* — In this section, we demonstrate more complicated cases where $\psi_\pm(\mathbf{r},z)$ are not simple eigen modes, that is, at least one of them is composed of $n$ eigen modes with different modal orders, leading to IPVs and IPRs occurring in both SOC space and spatial-mode subspace. We first consider a vector mode composed of three LG modes with different orders $N_1 = 1$, $N_2 = 5$, and $N_3 = 3$, respectively, and described by

$$\frac{1}{2}([LG_0^0(\mathbf{r},z)+LG_2^0(\mathbf{r},z)]\hat{e}_L + \frac{1}{\sqrt{2}}LG_0^1(\mathbf{r},z)\hat{e}_R \quad (8)$$
$$\delta^n(z) = \Delta N_n \arctan(z/z_R), \ n=1,2,3$$

Here $\delta^1(z)$ and $\delta^{2,3}(z)$ denote the IPVs which have accumulated in the spatial-mode subspace and the SOC space, respectively. Their corresponding modal-order difference are defined as $\Delta N_1 = N_2 - N_1 = 4$, $\Delta N_2 = N_3 - N_1 = 1$, and $\Delta N_3 = N_3 - N_2 = -3$, respectively, indicating IPR can only occur in the spatial-mode subspace at the far field. Therefore, as shown in Fig. 5(a), both the intensity and the polarization profiles of this mode change upon propagation, but only the intensity profile can be revived at $z_\infty$. We then consider another complex vector mode, given by

$$\frac{1}{2}[LG_1^2(\mathbf{r},z)+LG_3^2(\mathbf{r},z)]\hat{e}_L + \frac{1}{\sqrt{2}}LG_0^0(\mathbf{r},z)\hat{e}_R, \quad (9)$$

where the modal-order difference in the spatial-mode subspace is $\Delta N_1 = N_2 - N_1 = 4$ (a $2\pi$ IPV) while in the SOC space are $\Delta N_2 = N_1 - N_3 = 4$ (a $2\pi$ IPV) and $\Delta N_3 = N_2 - N_3 = 8$ (a $4\pi$ IPV). In contrast to Eq. (8), the IPRs in this mode can occur in both the spatial-mode subspace and the SOC space and, particularly, $\delta^{1,2,3}(z)$ would be synchronized at the far field. Figure 5(b) shows the evolution of $\delta^n(z)$ and the associated vector profiles upon propagation, and where the initial vector profile at the $z_0$ plane is revived at the far field ($z_\infty$). In the two above examples, it is important to note that because the variations and revivals of vector profiles involved both the intensity and polarization structures, thus the local polarization ellipticity, measured by concurrence, was changed upon propagation as well.

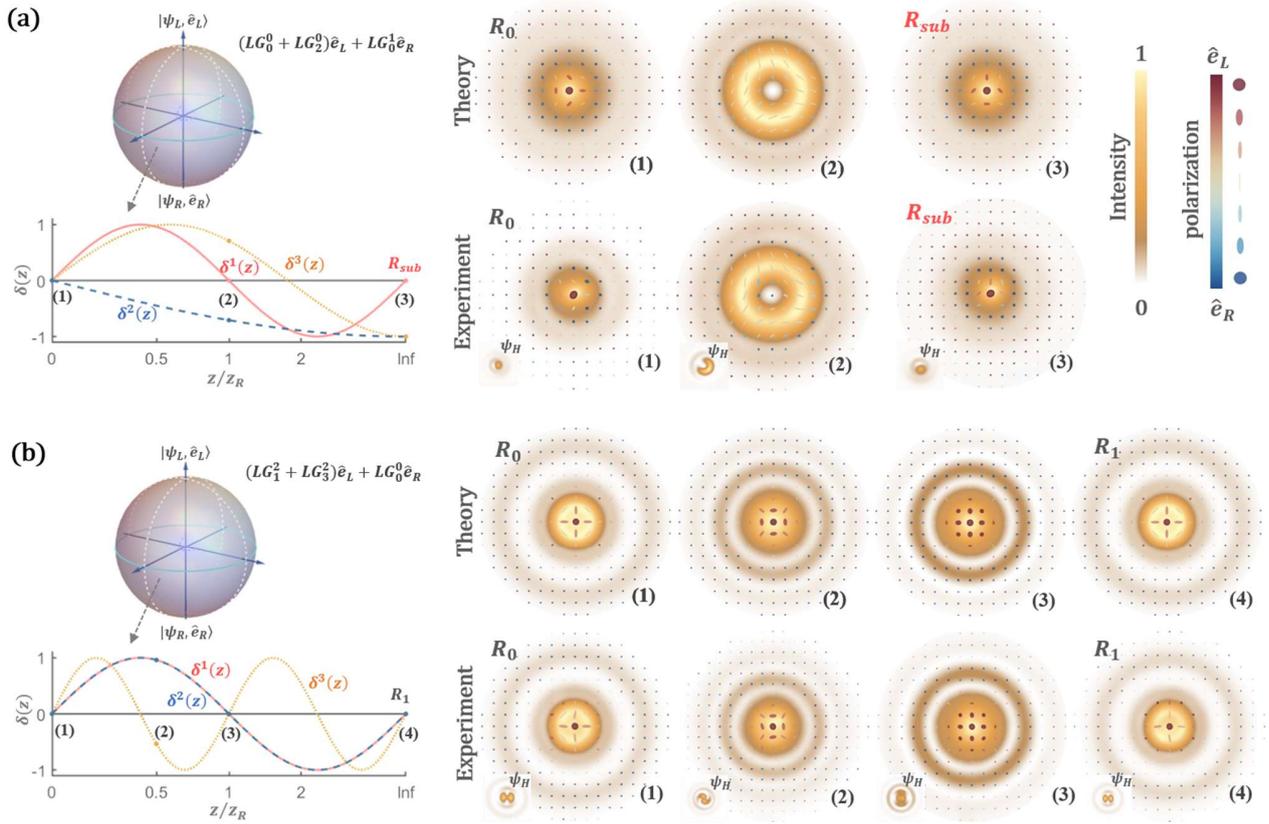

FIG. 5. Vector structured modes with propagation-variant vector profiles, where $R_n$ denotes nth. IPR and polarization-profile revival and $R_{\text{sub}}$ denotes IPR in spatial-mode subspace. The transverse unit is given in $r/w_z$.

Noteworthy, the evolution of vector profiles shown in Eqs. (8) and (9), in fact, is not rare. For instance, the FP modes used in most experiments were generated by using phase-only modulation, such as setting a q-plate as a 1/4 vortex waveplate by control voltage, in which the OAM carrying mode were not the eigen LG modes but the hyper-geometric Gaussian (HyGG) modes, and the corresponding FP modes can be expressed as

$$[\text{HyGG}^\ell(\mathbf{r},z)\hat{e}_L + LG_0^0(\mathbf{r},z)\hat{e}_R]/\sqrt{2}. \quad (10)$$

This HyGG mode carrying a well-defined OAM but undefined radial index, or rather, it can be represented as an infinite series of LG modes with same $\ell$ but different $p$ [45]. In consequence, the OAM carrying mode in Eq. (10) is not propagation invariant and the corresponding evolution of vector profile would be much stronger than those in Eqs. (8) and (9). In this case, it is impossible to predict easily the IPVs and IPRs upon propagation via using a concise toolkit shown in Eq. (4). But we still can obtain a numerical result by using diffraction integral, in specific, using Collins (or Kirchhoff) diffraction integral to predict the transverse structure at a given propagation plane, which in cylindrical coordinates is given by [4, 46]

$$E(r,\varphi,z) = \frac{i}{\lambda z}\exp(-ikz)\int r_0 dr_0 \int d\varphi_0 E(r_0,\varphi_0,0) \\ \exp\left\{-\frac{ik}{2z}\times[r_0^2 - 2rr_0\cos(\varphi-\varphi_0)+r^2]\right\}, \quad (11)$$

where the effective Gouy phase originates from the phase factor before the integral [29]. Figure 6 shows the theoretical and experimental results for the FP modes shown in Eq. (10) with $\ell = 1$ and $\ell = 2$, respectively. We see that, unlike the ideal FP modes based on LG modes, here the transverse plane only contains various linear polarizations at the $z_0$ plane, that is, '$C = 1$', while the well-known FP-type polarization profiles gradually appears on their way to the far field. Interestingly, for $\ell = 2$, a familiar cylindrical-vector-type polarization profile occurred at the beam waist plane.

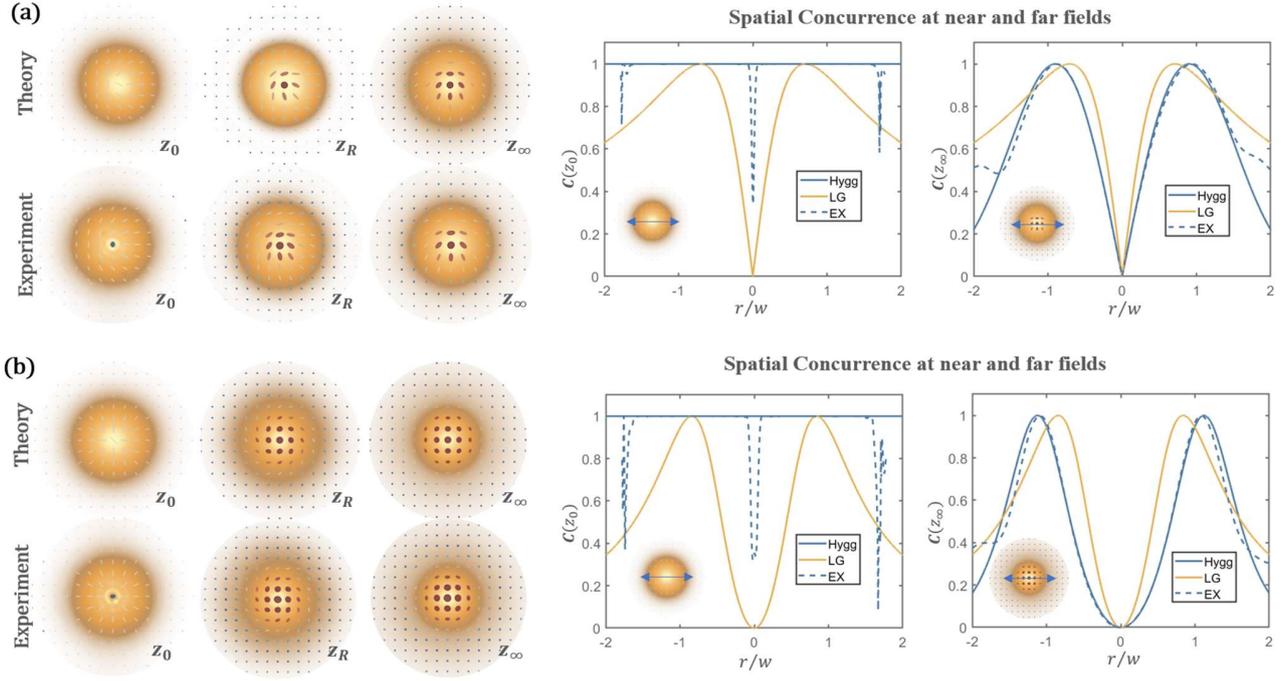

FIG. 6. Propagation evolution of FP modes based on HyGG modes with (a) $\ell = 1$ and (b) $\ell = 2$, respectively. The solid yellow lines in spatial concurrence corresponds to simulated results for LG-based FP modes, while solid and dash blue lines correspond to simulated and observed HyGG-based FP modes, respectively. The transverse unit is given in $r/w_z$.

## IV. Discussion & Conclusion

Based on above demonstrations, now it is clear that the propagation variations and revivals of the transverse structure presented above are governed by the IPVs and IPRs within non-eigen vector Gaussian modes. It can be intuitively understood the IPVs as the accumulated nonsynchronous Gouy phase during propagation, i.e., $\delta(z) = \Delta N\pi/2$ from $z_0$ to $z_\infty$. Furthermore, because the propagation of paraxial fields in free space or a lens system belong to a unitary transformation, and from this perspective, the Gouy-phase mediated IPVs and IPRs can therefore be regarded as a kind of Geometric phase effect. Figure 7 shows the 'motion path' (or parallel transport) of a spatial mode on the unitary sphere (or parameter space) of a 4f-lens transformation [31, 32]. Specifically, the red and blue traces denote propagations from $z_0$ to $z_\infty$ and vice versa, respectively. This motion encompasses a $2\pi$ solid angle on the sphere, and thus acquires an accompanied Geometric phase of $(N+1)\pi$. Namely, the fractional Gouy phase phenomenon, in fact, is the Geometric phase difference between the spatial modes with different modal orders under the same unitary transformation.

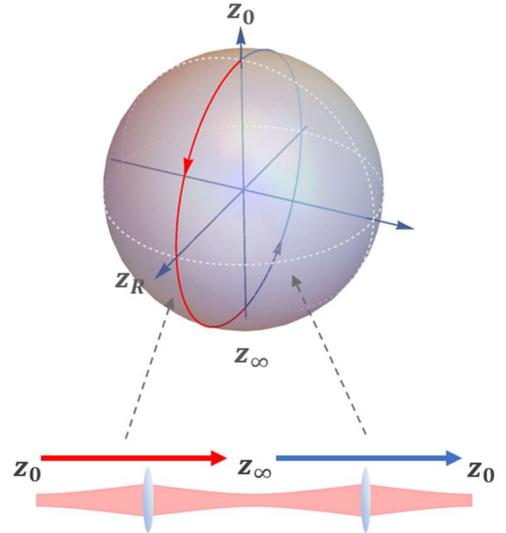

FIG. 7. Schematic of the geometric phase of spatial modes accompanying an 4f-imaging transformation.

In summary, we have systemically studied the evolution of transverse structure of non-eigen vector Gaussian mode upon propagation. The Gouy-phase mediated IPVs and IPRs in the non-eigen vector modes, as well as the associated influence on the 3D vector structure, are revealed. Our results provide a general principle for shaping vector structure light having propagation-invariant, on-demand variations, or periodic-revival structures, and the corresponding underlying principle may inspire many new photonic applications based on vectorially structured light.

## ACKNOWLEDGMENTS

Z.-H. Z., C. R.-G., and B.-S. S. acknowledge financial support from the National Natural Science Foundation of China (Grant Nos. 62075050, 11934013, and 61975047).

______________________________________